\documentclass[aps,prd,twocolumn,showpacs,floatfix,preprintnumbers,amsfont,amsmath,amssymb,nofootinbib,superscriptaddress]{revtex4-1}

\usepackage{color}
\usepackage{bbold}
\usepackage[colorlinks=true,linktocpage=true,linkcolor=colorLink,citecolor=colorCite,urlcolor=colorURL]{hyperref}
\usepackage[normalem]{ulem}
\usepackage{lipsum}
\usepackage{url}
\usepackage{float}
\usepackage[table]{xcolor}
\usepackage{graphicx}
\usepackage[font=small, justification=raggedright, format=plain]{caption}
\usepackage{subcaption}
\usepackage{mathtools}
\usepackage{bm}
\usepackage{siunitx}
\usepackage{multirow}
\definecolor{colorLink}{rgb}{0.9,0,0} 
\definecolor{colorCite}{rgb}{0,0.7,0} 
\definecolor{colorURL} {rgb}{0,0,0.8} 
\usepackage{fontawesome}

\usepackage{array,etoolbox}
\preto\tabular{\setcounter{magicrownumbers}{-1}}
\newcounter{magicrownumbers}

\colorlet{RED}{red}

\newcommand{\sk}[1]{}

\newcommand{\be}{\begin{equation}}
\newcommand{\ee}{\end{equation}}
\newcommand{\ba}{\begin{eqnarray}}
\newcommand{\ea}{\end{eqnarray}}

\newcommand{\apjl}{Astrophys. J. Lett.}

\newcommand{\mnras}{Mon. Not. R. Astron. Soc.}

\newcommand{\araa}{Annual Rev. Astron. Astrophys.}

\allowdisplaybreaks

\begin{document}

\title{Binary black hole population inference combining confident and marginal events \\from the $\tt{IAS\text{-}HM}$ search pipeline}

\author{Ajit Kumar Mehta}
\email{ajitkm@cmi.ac.in}
\affiliation{\mbox{Department of Physics, University of California at Santa Barbara, Santa Barbara, CA 93106, USA}}
\affiliation{\mbox{Chennai Mathematical Institute, Siruseri, India}}
\author{Digvijay Wadekar}
\affiliation{\mbox{Department of Physics and Astronomy, Johns Hopkins University,
3400 N. Charles Street, Baltimore, Maryland, 21218, USA}}
\affiliation{\mbox{School of Natural Sciences, Institute for Advanced Study, 1 Einstein Drive, Princeton, NJ 08540, USA}}
\author{Isha Anantpurkar}
\affiliation{\mbox{Department of Physics, University of California at Santa Barbara, Santa Barbara, CA 93106, USA}}
\author{Javier Roulet}
\affiliation{Theoretical AstroPhysics Including Relativity and Cosmology, California Institute of Technology, Pasadena, California, USA}
\author{Tejaswi Venumadhav}
\affiliation{\mbox{Department of Physics, University of California at Santa Barbara, Santa Barbara, CA 93106, USA}}
\affiliation{\mbox{International Centre for Theoretical Sciences, Tata Institute of Fundamental Research, Bangalore 560089, India}}
\author{Tousif Islam}
\affiliation{Kavli Institute for Theoretical Physics, University of California Santa Barbara, Santa Barbara, CA 93106}
\affiliation{Theoretical AstroPhysics Including Relativity and Cosmology, California Institute of Technology, Pasadena, California, USA}
\author{Jonathan Mushkin}
\affiliation{\mbox{Department of Particle Physics \& Astrophysics, Weizmann Institute of Science, Rehovot 76100, Israel}}
\author{Barak Zackay}
\affiliation{\mbox{Department of Particle Physics \& Astrophysics, Weizmann Institute of Science, Rehovot 76100, Israel}}
\author{Matias Zaldarriaga}
\affiliation{\mbox{School of Natural Sciences, Institute for Advanced Study, 1 Einstein Drive, Princeton, NJ 08540, USA}}
\date{\today}


\begin{abstract}

We present the population properties of binary black hole mergers identified by the $\tt{IAS\text{-}HM}$ pipeline (which incorporates higher-order modes in the search templates) during the third observing run (O3) of the LIGO, Virgo, and KAGRA (LVK) detectors. In our population inference analysis, instead of only using events above a sharp cut based on a particular detection threshold (e.g., false alarm rate), we use a Bayesian framework to consistently include both marginal and confident events. We find that our inference based solely on highly significant events ($p_{\mathrm{astro}} \sim 1$) is broadly consistent with the GWTC-3 population analysis performed by the LVK collaboration. However, incorporating marginal events into the analysis leads to a preference for stronger redshift evolution in the merger rate and an increased density of asymmetric mass-ratio mergers relative to the GWTC-3 analysis, while remaining within its allowed parameter ranges. Using simple parametric models to describe the binary black hole population, we estimate a merger rate density {[$R(z)=\mathcal{R}_0 (1+z)^\kappa$]} of $32.4^{+18.5}_{-12.2}\ \mathrm{Gpc}^{-3}\,\mathrm{yr}^{-1}$ at redshift $z = 0.2$, and a redshift evolution parameter of $\kappa = 4.4^{+1.9}_{-2.0}$. Assuming a power-law form for the mass ratio distribution ($\propto q^{\beta}$), we infer $\beta = 0.1^{+1.9}_{-1.4}$, indicating a relatively flat distribution. These results highlight the potential impact of marginal events on population inferences and motivate future analyses with data from upcoming observing runs. \href{https://zenodo.org/records/16893763}{\faDatabase}

\end{abstract}

\maketitle

\section{Introduction}
\label{sec:intro}

One of the main aims of gravitational-wave (GW) astronomy is to understand the population properties of the compact binary systems that produce these signals. Since the first detection in 2015 \cite{lvc_gw150914_firstGWdetection_prl2016abbott}, the LIGO, Virgo, and KAGRA (LVK) detectors have observed several tens of GW events by the end of the third observing run (O3) \cite{lvc_gwtc3_o3_ab_catalog_2021, nitz_4ogc_o3_ab_catalog_2021, Wad23_HM_Events}.\footnote{{We note that data from the first part of the fourth observing run (O4a) were released~\cite{LVKO4Data} during the preparation of this manuscript, increasing the number of confident detections to approximately 218.}} Although the number of detections is still in the Poisson-limited regime, it is now possible to begin constraining key features of the underlying astrophysical populations. Previous population studies have established several robust trends, including a bimodal structure in the primary mass distribution of binary black hole (BBH) mergers, evidence for redshift evolution in the merger rate, and indications of a preference for spin alignment with the orbital angular momentum \cite{LVKO3bpopulation, Boesky2024_DelayMetallicity, Fishbach2018_RateEvol, Tong2022, who_ordered_that-better_models_have_q_chieff_corr-Callister2021farr, RouletZaldarriaga2018, Farr2017}. As the catalog of BBH events continues to grow, we expect to uncover subtler features that will enable stringent tests of stellar evolution models, binary formation channels, and cosmological scenarios \cite{Galaudage2021, Vitale2019, Eldridge2016, Zevin_2021, Fishbach_2020, Map22_MergerRate, Van22_MergerRate, Boesky2024_DelayMetallicity,Bis22, Kus16_MergerTime_150914, chieff_isolated_field_binary_matias2018,Map19_MergerRate,Ger13_MassRatioReversal, Broekgaarden_2022_reversal, Pos14_mass_transfer,Rodriquez_2019_2g,Rod16_Dynamical}.

GW events are recovered from detector data using a variety of search pipelines \cite{PYCBCPipeline, Mes17_GstLAL_LowLatency, ias_pipeline_o1_catalog_new_search_prd2019, DalCanton:2017ala}, each implementing different assumptions and sometimes targeting different classes of signals. The LVK collaboration combines the results of five such pipelines to construct their gravitational-wave transient catalogs (GWTC) \cite{lvc_gwtc3_o3_ab_catalog_2021, GWTC-2.1}. Independent groups also perform searches with their own pipelines, producing complementary catalogs \cite{ias_pipeline_o1_catalog_new_search_prd2019, ias_o2_pipeline_new_events_prd2020, Ols22_ias_o3a, NitzCatalog_1-OGC_o1_2018, NitzCatalog_2-OGC_o2_2020, nitz_o3a_3ogc_catalog_2021, nitz_4ogc_o3_ab_catalog_2021,Che25_IMRI_search,Chi23,Meh23_ias_o3b, Kol24_Ares_ML_Search,Mar25_Aframe_ML}. The $\tt{IAS\text{-}HM}$ pipeline \cite{Wad23_Pipeline} is one such independent search pipeline. 

Nearly all the previous matched-filtering searches have included only the dominant quadrupole mode in the search templates, i.e., omitting higher-order harmonics which are predicted by general relativity. Ref.~\cite{Cha22} included higher-order modes (HMs) in the search templates, however their search was restricted to only near edge-on configurations. Ref.~\cite{Wad23_HM_Events} incorporated HM across the entire search parameter space by using the efficient mode-by-mode filtering implemented in the $\tt{IAS\text{-}HM}$ pipeline \cite{Wad23_Pipeline}. Running their pipeline on the LVK O3 data, Ref.~\cite{Wad23_HM_Events} produced a catalog that recovers most of the LVK GWTC-3 events, while adding $\sim 11$ new BBH candidates \cite{Wad23_HM_Events}. Subsequently, Ref.~\cite{Mehta:2025jiq} performed an injection-recovery campaign using the $\tt{IAS\text{-}HM}$ pipeline and found that including HMs in the search significantly increased the sensitive volume for high-redshift, high-mass and unequal-mass binaries. However, the new candidate events in the IAS-catalog were not high-significance detections; they have astrophysical probabilities $p_{\mathrm{astro}} < 1$ and inverse false alarm rates (IFARs) $\lesssim$ 0.7 yr. Nonetheless, they exhibit interesting trends such as higher redshifts, larger total masses, and more unequal mass ratios. 

As we will show in this work, these marginal events can still leave measurable imprints on the inferred BBH population. This happens because any individual marginal event comes with some probability of being of terrestrial origin, but observing a number of such events can carry substantial statistical information regarding their population. Moreover, if we do not consistently allow for the probability that at least some of the events in a population analysis could be of terrestrial origin, we run the risk of biasing the resulting inference. This underscores the importance of {consistently} including marginal events in population analyses \cite{Smith:2020lkj}.

The standard approach to inferring the population properties of GW sources is the hierarchical Bayesian framework \cite{Loredo2004, Mandel2019, Thrane2019, Loredo2013}, in which one specifies a prior model for the distribution of source parameters. This model may be parametric, astrophysically informed, or non-parametric. Each choice carries trade-offs: parametric or astrophysically motivated models can yield biased inferences if the assumed form is far from reality, while non-parametric models are more flexible, but typically harder to interpret and suffer from larger statistical uncertainties. In this work, we adopt simple parametric forms to describe the BBH mass and redshift distributions. Specifically, we use the standard $\tt{POWER\ LAW} + \tt{PEAK}$ model for the primary mass distribution, and simple power-law models for both the mass ratio and redshift evolution from Ref.~\cite{Talbot:2018cva,LVKO3bpopulation}. Spin distributions are not modeled here, as the inclusion of marginal events is not expected to qualitatively change existing spin-related results, given the comparatively low precision of spin measurements in GW parameter estimation; however, this is an avenue for future work.

Most population studies analyze a subset of candidate events imposing a somewhat high significance threshold, and assume in the likelihood calculation that they are all of astrophysical origin \cite{LVKO3bpopulation, Nitz:2021zwj, Wolfe:2025yxu}. In contrast, here we follow the formalism of Refs.~\cite{Roulet:2020wyq, Gae19_Pop_Subthreshold, Galaudage:2019jdx}, which allows the consistent inclusion of events with arbitrary $p_{\mathrm{astro}}$ values. This approach requires $p_{\mathrm{astro}}$ estimates computed with respect to a reference population model (the injection distribution), and reweights these values during the analysis according to the population model of interest. For our analysis, we use $\tt{IAS\text{-}HM}$ detections with inverse false-alarm rate (IFAR) $> 0.2$ yr, as further lowering the threshold does not change the conclusion of our results.

We find that restricting the analysis to high-IFAR events (where $p_{\mathrm{astro}} \to 1$) produces results well consistent with the LVK’s GWTC-3 population study. In contrast, incorporating marginal-significance events leads to noticeable shifts in the inferred population properties, most prominently an increase in the redshift evolution parameter and a higher inferred fraction of asymmetric mass-ratio mergers. This demonstrates that such lower-significance events can meaningfully influence population inferences. The inferred primary mass ($m_1$) distribution remains largely unchanged, still exhibiting two local peaks near $m_1 \sim 10\,M_{\odot}$ and $m_1 \sim 35\,M_{\odot}$, with no indication of an upper stellar-mass gap \cite{Farmer:2019jed,Mehta:2021fgz}. The secondary mass distribution also shows no clear evidence of such a gap, though its shape is less well constrained due to larger statistical uncertainties.

If BBHs formed exclusively from isolated binary stars undergoing common-envelope evolution, their merger rate [$R(z)=\mathcal{R}_0 (1+z)^\kappa$] should roughly trace the cosmic star formation rate (SFR), which at low redshift follows a power-law slope of $\kappa \sim 2-3$~\cite{Mad14_SFR}. Our population analysis yields $\kappa = 4.4^{+1.9}_{-2.0}$, noticeably steeper than the SFR trend. Although this slope is still statistically consistent with the SFR evolution within uncertainties, its high median value may indicate an enhanced contribution from alternative formation channels (such as primordial BBHs) at earlier cosmic times.  Dynamically assembled BBHs in young massive clusters could also produce a steeper evolution if cluster formation tracks the more rapid buildup of dense star-forming regions at high redshift. {Discriminating between these possibilities will require improved statistics from upcoming observational runs. Data from the first part of the fourth observing run (O4a) have recently been released \cite{LVKO4Data}, and we plan to present an updated analysis incorporating these data in forthcoming work. Notably, the recent GWTC-4 population results~\cite{LVKO4aPop} also indicate a slightly higher median value of the slope parameter ($\kappa \simeq 3.2$) compared to GWTC-3 ($\kappa \simeq 2.9$). }   Similarly, our finding of a mass-ratio distribution consistent with a flat slope departs from the expectation for isolated binary evolution, which typically favors equal-mass systems, and may point to a more diverse mixture of formation pathways.

\begin{figure*}[t]
    \centering
    \includegraphics[width=0.95\textwidth]{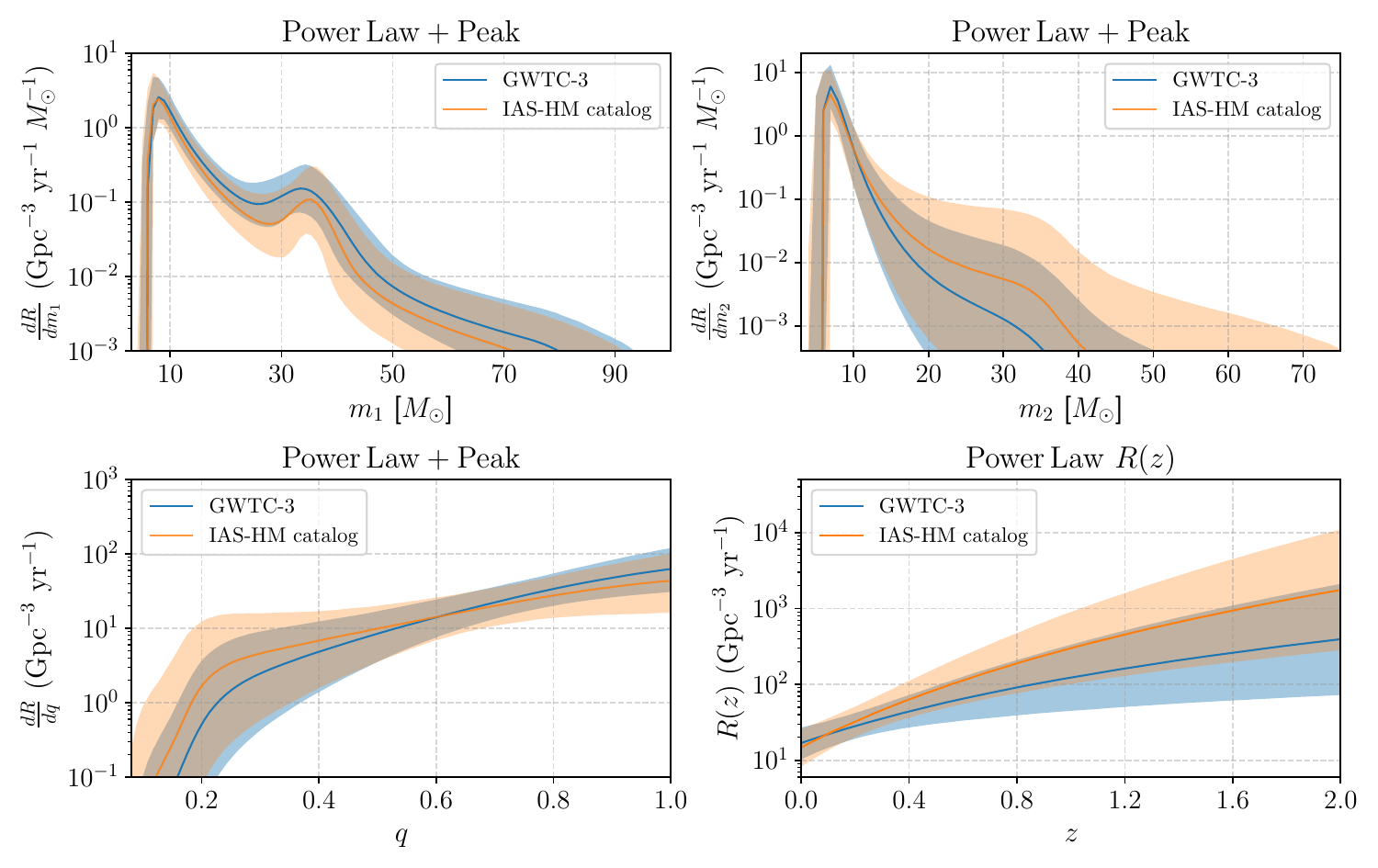}
    \vspace{-0.2cm}
\caption{\textit{Top panels}: Inferred differential merger-rate densities for primary mass $m_1$ and secondary mass $m_2$. We compare $\tt{IAS\text{-}HM}$ O3-catalog (including confident and marginal events with $\mathrm{IFAR}>0.2$ yr) to the GWTC-3 ($\mathrm{O1+O2+O3}$) population model. Medians and $90\%$ credible bands are shown. The $m_1$ distribution remains broadly consistent with GWTC-3, while the $m_2$ distribution is consistent within uncertainties but assigns slightly higher probability at the upper end of the spectrum.
\textit{Bottom panels}: Inferred differential merger-rate density as a function of mass ratio $q$ (left) and the redshift evolution $R(z)$ (right). The $q$ distribution is noticeably flatter than GWTC-3, favoring more asymmetric mergers. For $R(z)$, the $\tt{IAS\text{-}HM}$ inference favors a steeper rise with redshift, consistent with the larger growth rate ($\kappa$) obtained when marginal events are included.}
    \label{fig:m1_m2_q_post}
\end{figure*}

We structure this paper as follows. In Section~\ref{sec:method}, we outline the formalism for population inference and provide all relevant details needed to reproduce our analysis, including the datasets, prior assumptions, and estimates of the astrophysical probability ($p_{\mathrm{astro}}$). A detailed discussion of the $p_\mathrm{astro}$ computation is deferred to Appendix~\ref{app:p_astro}. Readers may wish to first consult Figure~\ref{fig:m1_m2_q_post}, which provides a visual overview of our key findings, before turning to the detailed discussion of results in Section~\ref{sec:results}. Section~\ref{sec:conclu} offers concluding remarks and directions for future research. \\

\section{Population likelihood}
\label{sec:method}

Consider a search pipeline that processes $T_{\rm obs}$ of LVK data and records $N_{\rm trig}$ triggers above a chosen detection-statistic threshold (e.g., IFAR). Let ${d_i}$ denote the data around the $i^\text{th}$ trigger ($i=1,2,\dots,N_{\rm trig}$), containing all information relevant to its classification. Assuming triggers are rare and independent, their number statistics would follow a Poisson distribution. Then, the likelihood of obtaining the dataset $(N_{\rm trig},\{d_i\})$ given a population model with hyper-parameters $\lambda$ describing the source population can be written as \cite{Roulet:2020wyq}
\begin{equation}
P(N_{\rm trig},\{d_i\}|\lambda) = P(N_{\rm trig}|\lambda) \prod_{i=1}^{N_{\rm trig}} P(d_i|\lambda),
\label{eq:lik}
\end{equation}
where {the Poisson probability is given by}
\begin{equation}
P(N_{\rm trig}|\lambda) = \frac{\left[N_a(\lambda) + N_b\right]^{N_{\rm trig}} e^{-\left[N_a(\lambda) + N_b\right]}}{N_{\rm trig}!}
\end{equation}
and {the normalized individual event probability is given by}
\begin{equation}
P(d_i|\lambda) = \frac{1}{N_a(\lambda) + N_b} \left[ \frac{\mathrm{d}N_a}{\mathrm{d}d}\bigg|_{d_i} (\lambda) + \frac{\mathrm{d}N_b}{\mathrm{d}d}\bigg|_{d_i} \right]
\end{equation}
Here, $N_a(\lambda)$ and $N_b$ are the expected numbers of astrophysical and background triggers, respectively, and $\mathrm{d}N_a/\mathrm{d}d$ and $\mathrm{d}N_b/\mathrm{d}d$ are their respective densities. {Notice that this is different from most of the previous population analysis, where it is assumed that only astrophysical triggers are present above a certain specified threshold (e.g., $\mathrm{IFAR} > 1$ yr in \cite{LVKO3bpopulation}) and possibility of having background triggers in the sample is neglected.}

Combining the above, Eq.~\eqref{eq:lik} can be written as
\begin{equation}
\begin{aligned}
P(N_{\rm trig},\{d_i\}|\lambda) &= \frac{e^{-N_a(\lambda) - N_b}}{N_{\rm trig}!} \prod_{i=1}^{N_{\rm trig}} \left[ \frac{\mathrm{d}N_a}{\mathrm{d}d}\bigg|_{d_i} (\lambda) + \frac{\mathrm{d}N_b}{\mathrm{d}d}\bigg|_{d_i} \right]
\end{aligned}
\label{eq:lik_final_0}
\end{equation}

For our implementation, it becomes convenient to re-normalize the likelihood by its value under a fixed reference model $\lambda_0$. This yields:
\begin{widetext}
\begin{align}
    P(N_{\rm{trig}},  \{d_{i}\} | \lambda) 
    & \propto \frac{P(N_{\rm{trig}},  \{d_{i}\} | \lambda)}{P(N_{\rm{trig}},  \{d_{i}\} | \lambda_0)}
      \propto e^{-N_a(\lambda)}\, 
      \prod_{i=1}^{N_{\rm{trig}}} 
      \frac{\mathrm{d}N_a(\lambda) + \mathrm{d}N_b}{\mathrm{d}N_a (\lambda_0) + \mathrm{d}N_b}\Bigg |_{d_i} \nonumber \\
    &= e^{-N_a(\lambda)}\, 
       \prod_{i=1}^{N_{\rm{trig}}} \Bigg[ 
       \frac{\mathrm{d}N_a(\lambda) }{\mathrm{d}N_a(\lambda_0)} \Bigg|_{d_i}\, 
       p_{\mathrm{astro}, i} (\lambda_0) 
       + \big(1 - p_{\mathrm{astro}, i} (\lambda_0)\big)  
       \Bigg] \label{eq:lik_final}
\end{align}
\end{widetext}

where the astrophysical probability for trigger $i$ is
\begin{equation}
p_{\mathrm{astro},i}(\lambda) = \frac{\mathrm{d}N_a(\lambda)}{\mathrm{d}N_a(\lambda) + \mathrm{d}N_b}\bigg|_{d_i}
\label{eq:p_astro}
\end{equation}
In Appendix~\ref{app:p_astro}, we describe the procedure of computing the reference $p_{\mathrm{astro}}$ of the triggers using simulated injections.

The expected astrophysical trigger density can be expressed in terms of the physical merger rate through,
\begin{equation}
\frac{\mathrm{d}N_a(\lambda)}{\mathrm{d}d}\bigg|_{d_i} = \int \mathrm{d}\theta \, P(d_i|\theta) \, \frac{\mathrm{d}N_a}{\mathrm{d}\theta}(\theta|\lambda),
\label{eq:foreground_rate_den}
\end{equation}
with
\begin{equation}
\frac{\mathrm{d}N_a}{\mathrm{d}\theta}(\theta|\lambda) = \mathcal{R}_0\, f(\theta|\lambda'),
\label{eq:phys_merg}
\end{equation}
where $\mathcal{R}_0$ is the local merger rate density and $f(\theta)$ is the normalized distribution of parameters $\theta$. Note that distance and arrival time lack a natural scale and hence they are not normalizable. We have used the notation $\lambda = \{\mathcal{R}_0, \lambda'\}$.

The expected number of detected astrophysical events can be expressed as,
\begin{equation}
N_a(\lambda) = \mathcal{R}_0\, \overline{VT}(\lambda'),
\end{equation}
where
\begin{equation}
\overline{VT}(\lambda') = \int \mathrm{d}\theta \, f(\theta|\lambda') \, p_{\mathrm{det}}(\theta)
\label{eq:VT}
\end{equation}
is the population-averaged sensitive volume--time, and $p_{\mathrm{det}}(\theta)$ is the detection probability for parameters $\theta$. Thus, $\overline{VT}$  depends on the search pipeline and the chosen detection threshold.

\subsection{Events and Data}
In this work, we evaluate Eq.~\eqref{eq:lik_final} using events detected by the $\tt{IAS\text{-}HM}$ pipeline \cite{Wad23_Pipeline} above an IFAR threshold of $0.2$ yr \cite{Wad23_HM_Events}.  We checked that lowering the threshold further has only a modest impact on the inferred population parameters, making this choice a reasonable balance between including marginal events and maintaining computational efficiency. To compute the quantities in Eqs.~\eqref{eq:phys_merg} and \eqref{eq:VT}, we perform Monte Carlo importance sampling over the parameter estimation (PE) posterior samples for each event. A comprehensive description of the $\overline{VT}$ calculation for the $\tt{IAS\text{-}HM}$ pipeline can be found in Ref. \cite{Mehta:2025jiq}. For $\tt{IAS\text{-}HM}$ events that are also present in GWTC-3 \cite{lvc_gwtc3_o3_ab_catalog_2021} we use the PE samples produced by the LVK and released on Zenodo \cite{zenodoLVKPE}. This ensures direct comparability between our results and the LVK results. For $\tt{IAS\text{-}HM}$ events not in GWTC-3, we perform our own PE  using our standard PE framework, $\tt{cogwheel}$ \cite{Roulet:2022kot}, with the same prior as used in GWTC-3.

\subsection{$\tt{CogwheelPop}$: Likelihood implementation and Sampling}
We make the $\tt{cogwheel}$ package a modular tool for hierarchical Bayesian inference of compact binary populations; we call it $\tt{CogwheelPop}$ \cite{IshaInPrep}.  It allows direct implementation of Eq.~\eqref{eq:lik_final} together with user-defined priors $\pi(\lambda|\mathcal{H})$ on the hyper-parameters $\lambda$ under a population hypothesis $\mathcal{H}$.

In this study, we adopt flat priors on all hyper-parameters (unless explicitly stated otherwise), reflecting minimal prior preference while maintaining comparability with the LVK GWTC-3 analysis \cite{LVKO3bpopulation}. The posterior distribution of the hyper-parameters $\lambda$ is then obtained through Bayes' theorem,
\begin{equation}
    P(\lambda|\{d_{i}\}, N_{\rm{trig}}, \mathcal{H}) \propto P(N_{\rm{trig}}, \{d_{i}\} | \lambda)\, \pi(\lambda|\mathcal{H})
\end{equation}
We explore this posterior using the $\tt{PyMultiNest}$ nested-sampling algorithm, running with $2048$ live points and a log-evidence tolerance of $0.25$. 

Below, we fix the population hypothesis ($\mathcal{H}$) to {the standard $\tt{POWER\ LAW} + \tt{PEAK}$ model (see Appendix~\ref{app:astro_bbh_model})} and assess the impact of including marginal events by varying the IFAR threshold for the \texttt{IAS\text{-}HM} catalog.
   
\begin{figure*}[t]
    \centering
    \includegraphics[width=0.9\textwidth]{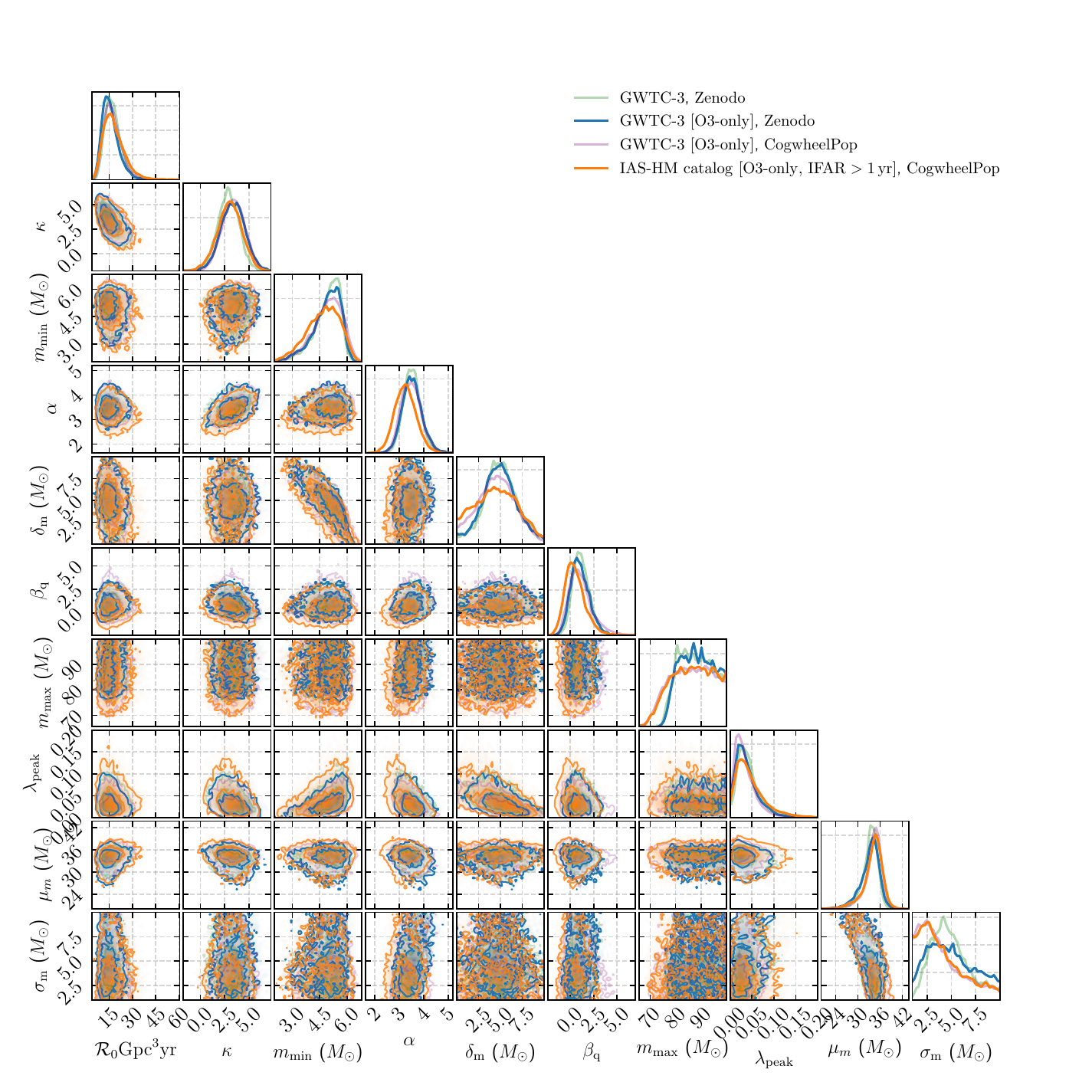}
    \vspace{-0.2cm}
    \caption{We first check that we obtain results consistent with the LVK analysis when we include only our higher significance ($\mathrm{IFAR} > 1$ yr) events in our analysis. We show posterior distributions of the $\tt{POWER\ LAW} + \tt{PEAK}$ hyperparameters for the GWTC-3 analysis (green), GWTC-3 O3-only analysis (blue), our $\tt{CogwheelPop}$ reproduction of the GWTC-3 O3-only result (purple), and the $\tt{IAS\text{-}HM}$ catalog with the same $\mathrm{IFAR} > 1$ yr threshold (orange).
    {Note that $\tt{CogwheelPop}$ \cite{IshaInPrep} is the population inference code used in our analysis (which is a generalized version of the 
    $\tt{cogwheel}$ \cite{Rou22_cogwheel} parameter estimation package for hierarchical Bayesian inference) and $\tt{Zenodo}$ corresponds to the hyperposterior samples taken from \cite{zenodoLVKPop}.}
    The close match between $\tt{IAS\text{-}HM}$ and GWTC-3 posteriors highlights the consistency between results of independent search pipelines.}
    \label{fig:Hyperparams_post_0}
\end{figure*}

\begin{figure*}[t]
    \centering
    \includegraphics[width=0.9\textwidth]{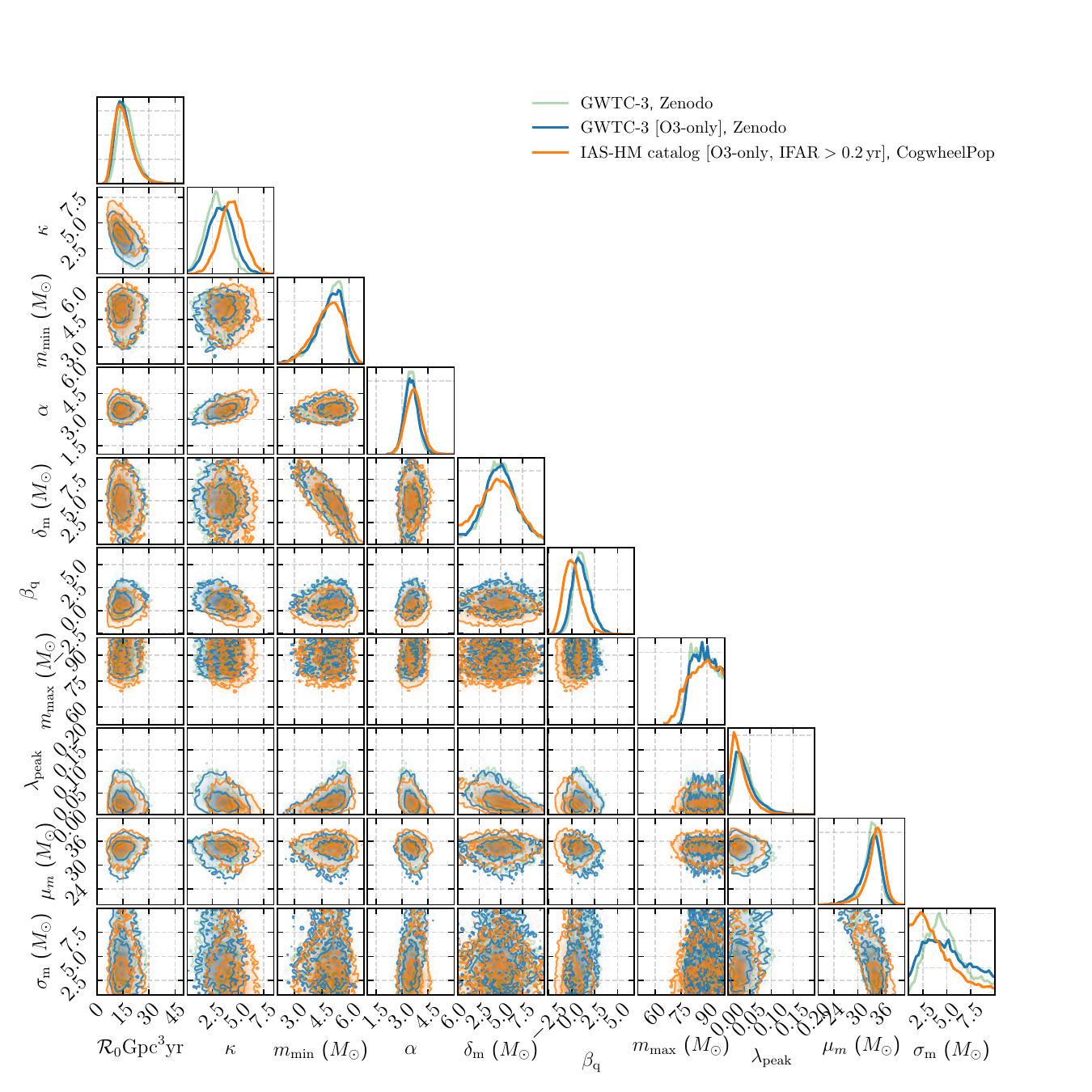}
    \vspace{-0.2cm}
    \caption{Similar to Fig.~\ref{fig:Hyperparams_post_0}, but also including marginal-significance events ($\mathrm{IFAR} \geq 0.2$ yr) from the $\tt{IAS\text{-}HM}$ catalog (orange) in our population inference analysis. Results are compared to the GWTC-3 (green) and GWTC-3 O3-only (blue) posteriors. The major differences upon including marginal events is that the redshift evolution parameter $\kappa$ increases and the mass-ratio slope $\beta_q$ is flatter.}
    \label{fig:Hyperparams_post_1}
\end{figure*}

\section{Discussion}
\label{sec:results}

\subsection{O3 LVK and $\tt{IAS\text{-}HM}$ events with $\mathrm{IFAR} > 1\, \rm{yr}$}

We begin by validating the $\tt{CogwheelPop}$ pipeline by reproducing the LVK's O3-only BBH population results, following the information provided in Ref.~\cite{LVKO3bpopulation}. Posterior samples for the O3 population analysis are taken from Zenodo~\cite{zenodoLVKPop}, along with the corresponding PE samples~\cite{zenodoLVKPE} and search sensitivity data~\cite{zenodoLVK}. These inputs are processed through $\tt{CogwheelPop}$ using exactly the same priors listed in Table~VI of Ref.~\cite{LVKO3bpopulation}. Since all GWTC-3 events considered here have $\mathrm{IFAR} > 1$ yr, we set $p_\mathrm{astro} = 1$ when evaluating the likelihood.

Figure~\ref{fig:Hyperparams_post_0} shows the hyperparameter posteriors for the $\tt{POWER\ LAW} + \tt{PEAK}$ BBH mass model. For reference, we show the GWTC-3 result from Zenodo (green) alongside the result obtained using only O3 data (blue), excluding O1 and O2 events. The similarity between these two curves indicates that adding O1+O2 data provides little additional constraining power for the parameters shown.

Our $\tt{CogwheelPop}$ reproduction of the O3-only GWTC-3 result is shown in purple. 
The agreement with the official LVK posteriors (blue or green) is very good, with only a noticeable difference in the width of the Gaussian peak ($\sigma_m$). 
Minor differences may arise from variations in sampling algorithms or other technical choices in the analysis codes, but we do not expect this to change our headline results which involve a comparison between runs of the same code, i.e., $\tt{CogwheelPop}$, with different event selection thresholds. 
We also note that our baseline spin model assumes isotropic spin directions and uniform spin magnitudes, whereas the LVK default spin model uses a mixture of isotropic and preferentially aligned tilts ($\theta_{1,2} \approx 0$) and the spin magnitudes drawn from a Beta distribution~\cite{LVKO3bpopulation}. When we adopt the LVK's spin prior, we find negligible changes in the hyperparameter posteriors shown in Figure~\ref{fig:Hyperparams_post_0}, so we use our baseline isotropic--uniform prescription for computational efficiency throughout this work.

The $\tt{IAS\text{-}HM}$ pipeline results for the same IFAR threshold ($\mathrm{IFAR} > 1$ yr) are shown in orange. The close agreement with the GWTC-3 posteriors (blue or green) is noteworthy, given that (other than analyzing the same dataset) $\tt{IAS\text{-}HM}$ is entirely independent of the five search pipelines used by the LVK for event identification and population inference. We note that the LVK analysis implicitly assumes that $\mathrm{IFAR} > 1$ yr corresponds to $p_{\mathrm{astro}} \to 1$, an assumption that may not hold for all events~\cite{lvc_gwtc3_o3_ab_catalog_2021}. In principle, this could introduce biases into the inferred population properties. However, the strong consistency with our analysis, which explicitly incorporates $p_\mathrm{astro}$ into the likelihood, suggests that such biases are negligible for this dataset.

\subsection{$\tt{IAS\text{-}HM}$ events with $\mathrm{IFAR} > 0.2\, \rm{yr}$}

We relax the IFAR threshold to $0.2$ yr, thereby incorporating lower-significance (marginal) events. The resulting population constraints are shown in Figure~\ref{fig:Hyperparams_post_1} (orange). As noted earlier, we adopt the same hyperparameter priors as those used in the GWTC-3 analysis \cite{LVKO3bpopulation}. In GWTC-3, the prior on $m_{\rm max}$ was restricted to values below $100\,M_{\odot}$, since not all LVK search pipelines report sensitivity estimates beyond this range. In contrast, the $\tt{IAS\text{-}HM}$ pipeline provides $\overline{VT}$ estimates up to $m_{\rm max} \lesssim 500\,M_{\odot}$. We therefore also performed a population analysis allowing $m_{\rm max} \leq 500\,M_{\odot}$. We find that the broader prior, however, has negligible impact on the results and we omit this result from Figure \ref{fig:Hyperparams_post_1} for brevity. 

{Note that the IAS-HM catalog contains a number of marginal events which have support for $m_1\gtrsim 100\,M_{\odot}$ \cite{Wad23_HM_Events}. We found that these events do not significantly affect the standard $\tt{POWER\ LAW} + \tt{PEAK}$ analysis. One could however perform a separate analysis by adding an extra component to the mass model which accounts for a potential IMBH (intermediate mass black hole) population. We leave this to a future paper.}
 
Comparing the orange curves to the GWTC-3 (green) or GWTC-3 O3-only (blue) results shows that including marginal events increases the redshift evolution parameter $\kappa$ by $\sim 1$ and reduces the mass-ratio slope $\beta_q$, indicating a flatter $q$ distribution. Figure~\ref{fig:m1_m2_q_post} shows the inferred differential merger rates for the primary mass and secondary mass in the top panel, and mass ratio and the redshift evolution in the bottom panel. The primary mass distribution remains broadly consistent with LVK’s GWTC-3 result, but the mass-ratio distribution is noticeably flatter---predicting a higher fraction of asymmetric BBH mergers. The secondary mass distribution is also broadly consistent within uncertainties, though at the higher-mass end it may favor slightly larger probabilities. The inferred redshift distribution (bottom-right panel) clearly shows that the $\tt{IAS\text{-}HM}$ catalog prefers higher merger rates at higher redshifts.

\begin{figure}[H]
    \centering
    \includegraphics[width=0.48\textwidth]{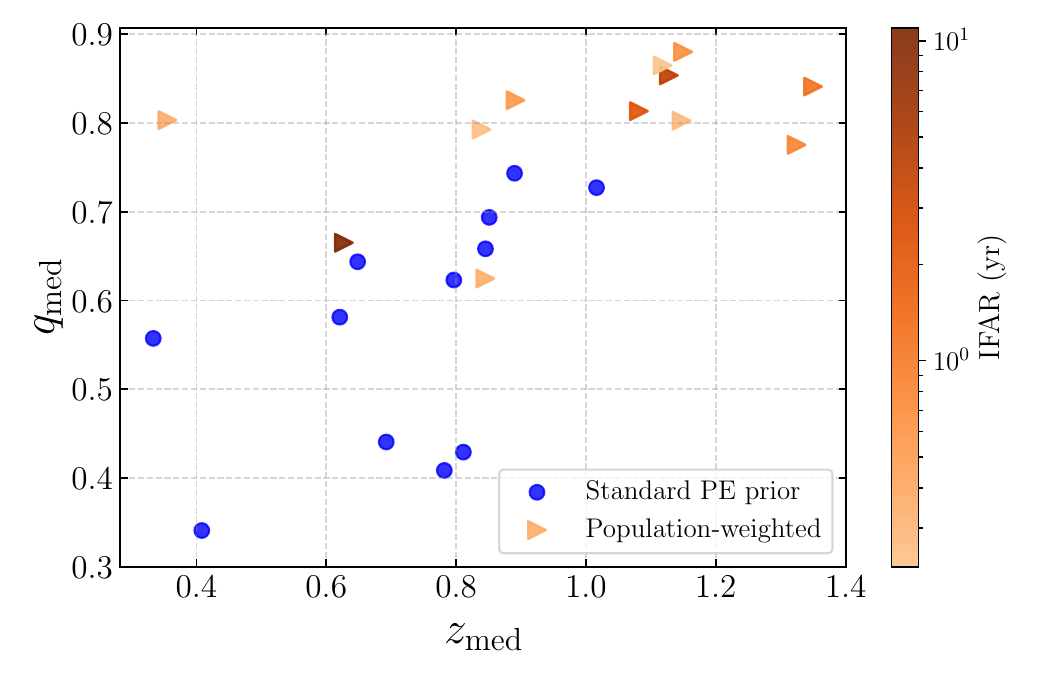}
    \vspace{-0.2cm}
\caption{Median posterior estimates of mass ratio $q$ and redshift $z$ for the $\tt{IAS\text{-}HM}$ {events with $\mathrm{IFAR}>0.2$\,yr which are not in} GWTC-3~\cite{lvc_gwtc3_o3_ab_catalog_2021}. Circular markers show medians under the standard LVK PE prior. Triangular markers show medians after reweighting with our population model and marginalizing over hyperparameter uncertainties, with their color 
    gradient indicating the IFAR values. Shifts toward higher $q$ and $z$ after reweighting reflect the coupling between the inferred $\kappa$ and $\beta_q$, as evident in Figure~\ref{fig:Hyperparams_post_1}.}
    \label{fig:q_vs_z}
\end{figure}

To investigate the origin of these differences, in Figure~\ref{fig:q_vs_z}, we plot the median PE values of $q$ and $z$ for the $\tt{IAS\text{-}HM}$ events which do not overlap with GWTC-3~\cite{lvc_gwtc3_o3_ab_catalog_2021}. We show both the standard LVK PE prior medians and the values obtained after reweighting the samples using our population results, marginalizing over hyperparameter uncertainties. The standard PE priors tend to yield lower $q$ and $z$ values, which are shifted toward higher $q$ and $z$ after population reweighting. This reflects the degeneracy between $\beta_q$ and $\kappa$, as also evident in Figure~\ref{fig:Hyperparams_post_1}. Notably, a few events, including one with a high IFAR ($\sim 11$ yr), retain median $q \lesssim 0.7$ even after population reweighting. These events are likely driving the likelihood toward larger values of $\beta_q$.

Thus, the steep redshift evolution we infer 
(\(\kappa \approx 4\)) may be partially driven by a degeneracy between the 
redshift evolution parameter \(\kappa\) and the mass-ratio slope \(\beta_q\). 
In our analysis, allowing \(\beta_q\) to take lower values (implying a higher 
fraction of asymmetric mergers) reduces the average detectability of BBHs, 
especially at high redshift, due to the lower signal-to-noise ratios of 
unequal-mass systems. 
To match the observed number of high-\(z\) detections, the model can compensate 
by increasing \(\kappa\), even if the underlying merger-rate evolution is not 
intrinsically that steep. 
This parameter correlation means that \(\kappa\) should not be over-interpreted 
as a direct measurement of the astrophysical merger-rate evolution. 
Breaking this degeneracy will require improved statistics from future observing 
runs or tighter external constraints on \(\beta_q\) from theoretical modeling, and careful joint modeling of the \(z\)--\(q\) 
distribution. We leave this to future work.

\section{Conclusion}
\label{sec:conclu}

In this work, we have performed a population analysis of BBH mergers detected by the $\tt{IAS\text{-}HM}$ search pipeline during the LVK O3 run. By explicitly incorporating events across a range of astrophysical probabilities ($p_\mathrm{astro}$), we have demonstrated the impact of marginal-significance candidates on inferred population properties. We first validated our new population inference framework ($\tt{CogwheelPop}$) by reproducing the LVK population results on their dataset, and then showed that the $\tt{IAS\text{-}HM}$ catalog yields consistent results with the LVK catalog when restricted to high-confidence detections ($\mathrm{IFAR}>1$ yr). This demonstrates that two independent search pipelines, $\tt{IAS\text{-}HM}$ and the suite of LVK pipelines, lead to mutually consistent population inferences when applied to the same underlying data, underscoring the robustness of current BBH population measurements. However, when marginal $\tt{IAS\text{-}HM}$ events are included self-consistently in our Bayesian analysis, we find a steeper redshift evolution of the merger rate and a flatter mass-ratio distribution, pointing to an enhanced fraction of asymmetric binaries.

These trends, while subject to parameter degeneracies, suggest that marginal events, often excluded from standard analyses, can contain valuable information about the underlying population. Their inclusion emphasizes the importance of accounting for both confident and lower-significance detections in order to obtain an unbiased view of the astrophysical population. Breaking the observed degeneracy between the redshift evolution parameter and the mass-ratio slope will require improved statistics from upcoming observing runs or prior information from theoretical population models.

Looking ahead, the rapid growth of the GW catalog in O4 and beyond will offer unprecedented opportunities to refine these measurements. In particular, extending the analysis framework to incorporate spin distributions, more flexible mass models, and possible correlations between source parameters will enable more stringent tests of binary formation scenarios and cosmological evolution. Our results highlight the utility of independent search pipelines and Bayesian treatments of marginal events as key ingredients in extracting the full astrophysical potential of GW observations.

To enable reproducibility of our work, we provide two separate Zenodo repositories. The first \cite{zenodoIAS} contains the O3 injection summary files required for computing $\overline{VT}$ of the $\tt{IAS\text{-}HM}$ pipeline. The second \cite{zenodoIASPop} contains the posterior samples of all events with $\mathrm{IFAR}>0.2$ yr, together with their IFARs and reference $p_{\rm astro}$ values.

\section*{Acknowledgements}

We thank Soumendra Kishore Roy, Sumit Kumar, Colm Talbot and Matthew Mould for helpful discussions.
TV acknowledges support from NSF grants 2012086 and 2309360, the Alfred P. Sloan Foundation through grant number FG-2023-20470, the BSF through award number 2022136 and the Hellman Family Faculty Fellowship. BZ is supported by the Israel Science Foundation, NSF-BSF and by a research grant from the Willner Family Leadership Institute for the Weizmann Institute of Science. MZ is supported by NSF 2209991, NSF-BSF 2207583, the Nelson Center for Collaborative Research and the Simons Foundation (Simons Collaboration on Black Holes and Strong Gravity). This research was also supported in part by the National Science Foundation under Grant No. NSF PHY-1748958.  

This research has made use of data, software and/or web tools obtained from the Gravitational Wave Open Science Center (\url{https://www.gw-openscience.org/}), a service of LIGO Laboratory, the LIGO Scientific Collaboration and the Virgo Collaboration. LIGO Laboratory and Advanced LIGO are funded by the United States National Science Foundation (NSF) as well as the Science and Technology Facilities Council (STFC) of the United Kingdom, the Max-Planck-Society (MPS), and the State of Niedersachsen/Germany for support of the construction of Advanced LIGO and construction and operation of the GEO600 detector. Additional support for Advanced LIGO was provided by the Australian Research Council. Virgo is funded, through the European Gravitational Observatory (EGO), by the French Centre National de Recherche Scientifique (CNRS), the Italian Istituto Nazionale di Fisica Nucleare (INFN) and the Dutch Nikhef, with contributions by institutions from Belgium, Germany, Greece, Hungary, Ireland, Japan, Monaco, Poland, Portugal, Spain.

\appendix
\section{Computation of the reference $p_{\rm{astro}}$ of the triggers}
\label{app:p_astro}

We compute the reference $p_{\rm astro}$ values for our triggers following the procedure outlined in Appendix~B of Ref.~\cite{Roulet:2020wyq}, with some modifications to account for our choice of reference astrophysical model and the inclusion of HMs in the $\tt{IAS\text{-}HM}$ pipeline. For completeness, we summarize the main steps of the procedure below.

As defined in Eq.~\ref{eq:p_astro}, the $p_{\mathrm{astro}, i}$ of a trigger $i$ depends on the local ratio of the foreground and background trigger rate densities at the corresponding data point $d_i$. While the foreground rate density can be computed directly (Eq.~\ref{eq:foreground_rate_den}), the background rate density cannot be obtained in an exact closed form. Instead, we adopt the following approximation, which we expect to be accurate for our purposes:
\begin{equation}
    \frac{\mathrm{d}N_a/\mathrm{d}d}{\mathrm{d}N_b/\mathrm{d}d} \Bigg|_{d_i} (\lambda_0)
    \approx
    \frac{\mathrm{d}N_a/\mathrm{d}\tilde{\rho}^2}{\mathrm{d}N_b/\mathrm{d}\tilde{\rho}^2}
    \big( \tilde{\rho}^2_i \,|\, \mathcal{T} \approx \mathcal{T}_i, \lambda_0 \big),
    \label{eq:dnabydnb}
\end{equation}
where $\tilde{\rho}^2$ is the $\tt{IAS\text{-}HM}$ detection statistic~\cite{Wad23_Pipeline} and $\mathcal{T}$ denotes the template parameters associated with a trigger. Since the detection statistic contains only coarse information about the source parameters, we explicitly condition on $\mathcal{T}$ to enhance discrimination between foreground and background triggers.

We evaluate Eq.~\ref{eq:dnabydnb} as follows. We generate a large set of \emph{foreground} triggers from injections drawn from the reference astrophysical model described by the hyperparameters $\lambda_0$ (specified below). The $\tt{IAS\text{-}HM}$ pipeline also provides a large set of \emph{background} triggers from time-slid analyses of the O3 LVK data~\cite{Wad23_HM_Events}. For each trigger $i$ with template $\mathcal{T}_i$, we select all foreground and background triggers with similar templates, requiring a match $\langle \mathcal{T}, \mathcal{T}_i \rangle$ above a chosen threshold. For simplicity, we compute the match using only the $(2,2)$ mode templates. We select the highest match threshold that retains at least $100$ injection triggers and $100$ background triggers, balancing locality in parameter space against statistical uncertainty.

For the selected triggers, we perform a kernel density estimate (KDE) of $\mathrm{d}N_a/\mathrm{d}\tilde{\rho}^2$ and $\mathrm{d}N_b/\mathrm{d}\tilde{\rho}^2$ using weights:
\begin{align}
    w_j^{\mathrm{inj}} &= \frac{\mathcal{R}_0^{\rm inj}\, V_c^{\rm inj}\, T_{\mathrm{obs}}}{N_{\mathrm{inj}}} \\
    w_j^{\mathrm{bg}}  &= \frac{1}{N_{\mathrm{timeslides}}}\\ \nonumber
\end{align}
for injections and background triggers, respectively. Here, $V_c^{\rm inj}$ is the total comoving volume accessible to the reference astrophysical model, $T_{\mathrm{obs}}$ is the observing time, and $N_{\mathrm{inj}}$ is the total number of injections in the set.

For the reference astrophysical model, we use the public LVK injection sets available on Zenodo~\cite{zenodoLVK}. The LVK provides four injection sets corresponding to binary neutron stars (BNS), neutron star–black hole systems (NSBH), BBH, and intermediate-mass black holes (IMBH). Since $\tt{IAS\text{-}HM}$ targets only the BH parameter space, we use the BBH and IMBH injection sets.

\begin{figure}[t]
    \centering
    \includegraphics[width=0.47\textwidth]{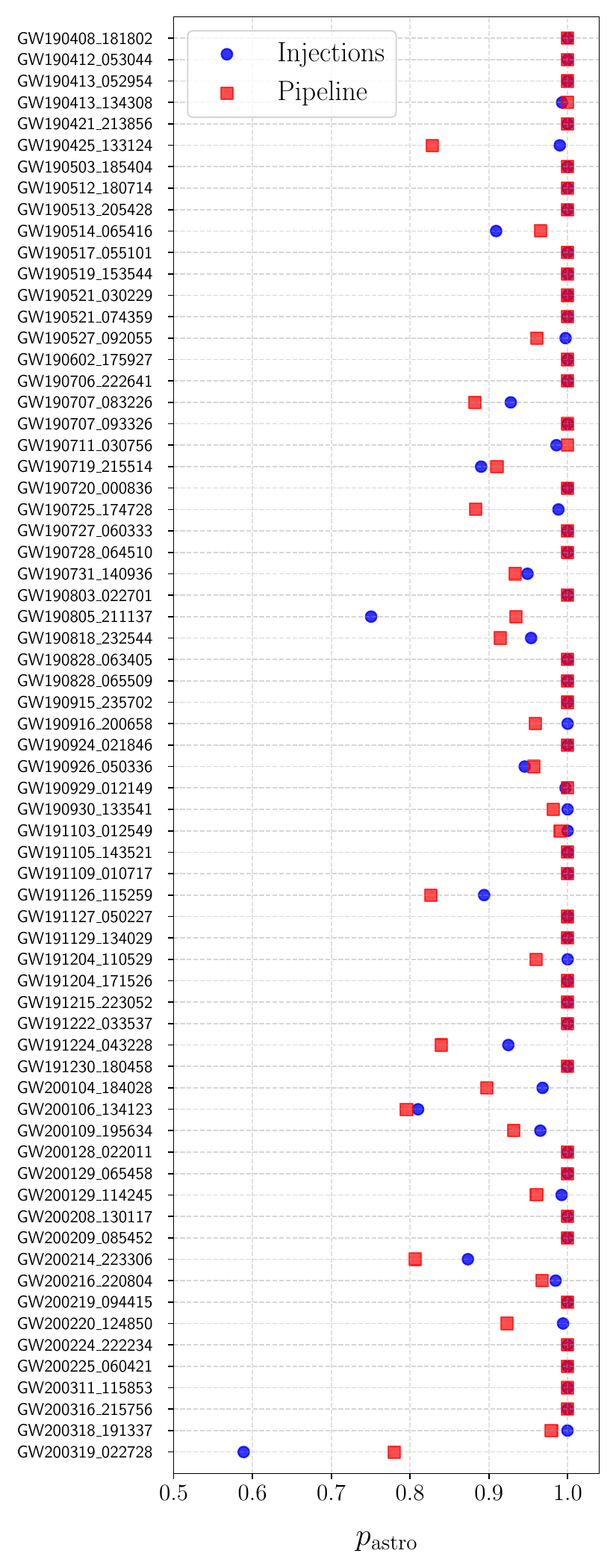}
    \vspace{-0.4cm}
    \caption{Astrophysical probabilities, $p_{\mathrm{astro}}$, for events (ordered here by their trigger time, $t_\mathrm{gps}$) with $\mathrm{IFAR} > 0.2$ yr ($x$-axis), computed using the injection set described in Appendix~\ref{app:p_astro}. For comparison, we also show the $p_{\mathrm{astro}}$ values assigned directly by the $\tt{IAS\text{-}HM}$ pipeline, which estimates them by comparing the ranking score densities of coincident triggers between the Livingston and Hanford detectors with those of other (i.e., background) triggers.
    }
    \label{fig:pastro_trigs}
\end{figure}

We compute $\mathrm{d}N_a / \mathrm{d}N_b$ separately for the BBH and IMBH sets, and then combine them to compute $p_{\rm{astro}}$ using Eq.~\ref{eq:p_astro} :
\begin{equation}
    \frac{\mathrm{d}N_a}{\mathrm{d}N_b} \Bigg|_{d_i}
    =
    \frac{\mathrm{d}N_a^{\rm BBH}}{\mathrm{d}N_b} \Bigg|_{d_i}
    +
    \frac{\mathrm{d}N_a^{\rm IMBH}}{\mathrm{d}N_b} \Bigg|_{d_i}.
    \label{Eq:dna_dnb}
\end{equation} \\

For computing the weights, we obtain $V_c^{\rm inj}$ by integrating the differential comoving volume $\mathrm{d}V_c/\mathrm{d}z$ up to the maximum redshift of each injection set:
\begin{align}
    V_c^{\mathrm{inj,\, BBH}}  &= 575.38 \, \mathrm{Gpc}^{3}, \\
    V_c^{\mathrm{inj,\, IMBH}} &= 365.14 \, \mathrm{Gpc}^{3}.
\end{align}

To set the rate normalizations, we resort to the $\tt{IAS\text{-}HM}$ pipeline. The pipeline detects $67$ events above an IFAR threshold of $0.2$ yr. For the BBH injection set, defined with primary masses in the range $2\,M_{\odot} \leq m_1 \leq 100\,M_{\odot}$ and a power-law slope $\alpha = -2.35$, we obtain an average sensitive volume–time of $\overline{VT} \sim 0.32 \, \mathrm{Gpc^3\, yr}$ \cite{Mehta:2025jiq}. For the IMBH injection set, we adopt a rough normalization. Accordingly, we make the following choices:
\begin{align}
    \mathcal{R}_0^{\rm inj,\, BBH}  &= 220 \, \mathrm{Gpc}^{-3}\,\mathrm{yr}^{-1}, \\
    \mathcal{R}_0^{\rm inj,\, IMBH} &=  50 \, \mathrm{Gpc}^{-3}\,\mathrm{yr}^{-1}.
\end{align}
The absolute values of these rates do not affect the population analysis results, provided they are used consistently for $p_{\mathrm{astro}}$ computation and in the likelihood expression (Eq.~\eqref{eq:lik_final}). The relatively high BBH rate reflects the LVK’s choice of an injection mass range overlapping with neutron-star masses to facilitate the construction of mixed-source injection sets.

In Figure~\ref{fig:pastro_trigs}, we show the reference $p_{\rm{astro}}$ values for triggers with $\mathrm{IFAR}>0.2$ yr computed using the injections, along with the pipeline’s $p_{\rm astro}$. 
The two calculations differ in the choice of astrophysical model, including the distribution of source parameters and the merger rate. The pipeline result marginalizes over the rate and assumes: ($i$) a simple power law prior for the detector frame total masses $P(M^\mathrm{det}_\mathrm{tot})\propto M_\mathrm{tot}^{-2}$ ($ii$) uniform distribution in $q$ $(iii)$ flat in $\chi_\mathrm{eff}$, $(iv)$ flat in luminosity volume distribution.
The pipeline $p_\textrm{astro}$ is computed following \cite{Wad23_HM_Events}, and the injection $p_\textrm{astro}$ following \cite{Roulet:2020wyq}.

\section{$\tt{POWER\ LAW} + \tt{PEAK}$ model of BBH Parameters}
\label{app:astro_bbh_model}

\subsection{Primary mass ($m_1$) and mass ratio ($q$)}
In this model (for Eq.~\ref{eq:phys_merg}), we assume that the primary BH mass $m_1$ follows a mixture distribution composed of a power law and a Gaussian component (see also Appendix~B of Ref.~\cite{LVKO3bpopulation}):
\begin{align}
    \pi(&m_1 \mid \alpha, m_{\min}, \delta_m, m_{\max}, 
    \lambda_{\rm{peak}}, \mu_m, \sigma_m) \nonumber \\
    &= \Big[ (1-\lambda_{\rm{peak}})\, \mathcal{P}(m_1 \mid -\alpha, m_{\max}) \nonumber \\
    &\quad + \lambda_{\rm{peak}}\, \mathcal{G}(m_1 \mid \mu_m, \sigma_m) \Big] \,
     S(m_1 \mid m_{\min}, \delta_m)
\end{align}

Here, $\mathcal{P}(m_1 \mid -\alpha, m_{\max})$ is a normalized power-law distribution with slope $-\alpha$ and an upper cutoff at $m_{\max}$, while $\mathcal{G}(m_1 \mid \mu_m, \sigma_m)$ is a normalized Gaussian distribution with mean $\mu_m$ and standard deviation $\sigma_m$. The mixing fraction $\lambda_{\rm{peak}}$ controls the relative weight of the Gaussian “peak” component. The entire mixture is multiplied by a tapering function $S(m_1 \mid m_{\min}, \delta_m)$, which smoothly suppresses the distribution below $m_{\min}$ by transitioning from $0$ to $1$ across the interval $[m_{\min},\, m_{\min} + \delta_m]$.

The tapering function is defined as
\begin{align}
S& (m \mid m_{\min}, \delta_m) = \nonumber\\[-4pt]
& \left\{
\begin{array}{ll}
0, & m < m_{\min}, \\[6pt]
\bigl[f(m - m_{\min}, \delta_m) + 1\bigr]^{-1}, 
& m_{\min} \le m < m_{\min} + \delta_m, \\[10pt]
1, & m \ge m_{\min} + \delta_m,
\end{array}
\right.
\tag{B5}
\end{align}
with
\begin{align}
f& (m', \delta_m) =
\exp\!\left(\frac{\delta_m}{m'} + \frac{\delta_m}{m' - \delta_m}\right)
\tag{B6}
\end{align}

For the mass ratio $q = m_2/m_1$, we adopt a power-law distribution with slope $\beta_q$:  
\begin{equation}
    \pi(q \mid \beta_q, m_1, m_{\min}, \delta_m) \propto 
    q^{\beta_q}\, S(qm_1 \mid m_{\min}, \delta_m)
\end{equation}
The tapering function $S$ is again applied to ensure the physical requirement that both component masses exceed $m_{\min}$.

\subsection{Spin model}
For each BH spin vector ($\boldsymbol{\chi}_{1,2}$), we assume an isotropic distribution in direction and a uniform distribution in magnitude between $0$ and a maximum value $\chi_{\max}$:
\begin{equation}
\pi(\boldsymbol{\chi}_i) =
\frac{1}{4\pi\, |\boldsymbol{\chi}|^{2}\, \chi_{\max}}\,
\Theta\!\left(|\boldsymbol{\chi}| \leq \chi_{\max}\right),
\end{equation}
where $\Theta$ is the Heaviside step function and $i \in \{1,2\}$. We take $\chi_{\max}=0.998$.  
We also tested the default spin model adopted in the GWTC-3 population analysis~\cite{LVKO3bpopulation} and found no significant impact on the posteriors of the mass and redshift-evolution hyperparameters considered in this work.

\subsection{Merger rate density $R(z)$}
The redshift evolution of the BBH merger rate density is modeled as a power law:
\begin{equation}
    R(z) = \mathcal{R}_0 \,(1+z)^{\kappa},
\end{equation}
where $\mathcal{R}_0$ is the local merger rate density and $\kappa$ is the growth index.

\subsection{Priors for the hyperparameters}
In Table~\ref{tab:PLPeak_params_prior}, we summarize the priors adopted for the hyperparameters of the $\tt{POWER\ LAW + PEAK}$ model. All parameters are assigned uniform priors, except for $\mathcal{R}_0$, which follows a log-uniform distribution. For the $\tt{IAS\text{-}HM}$ analysis including low-IFAR events ($>0.2$ yr), we extend the prior ranges only for $\beta$ and $\kappa$, to $\beta \in [-6,7]$ and $\kappa \in [-10,15]$, as required by the likelihood support. 

\begin{table*}[t]
\centering
\caption{Priors for $\tt{POWER\ LAW + PEAK}$ model parameters.}
\begin{tabular}{lp{0.7\textwidth}l}
\hline\hline
Parameter & Description & Prior \\
\hline
$\mathcal{R}_0$ & Local BBH merger rate density & $\mathrm{LogU}(1,\,200)$ \\
$\alpha$       & Spectral index for the power law of the primary mass distribution & $U(-4,\,12)$ \\
$\beta_q$      & Spectral index for the power law of the mass ratio distribution   & $U(-2,\,7)$ \\
$m_{\min}$     & Minimum mass of the power-law component of the primary mass distribution & $U(2M_\odot,\,10M_\odot)$ \\
$m_{\max}$     & Maximum mass of the power-law component of the primary mass distribution & $U(50M_\odot,\,100M_\odot)$ \\
$\lambda_{\rm peak}$ & Fraction of BBH systems in the Gaussian component & $U(0,\,1)$ \\
$\mu_m$        & Mean of the Gaussian component in the primary mass distribution & $U(20M_\odot,\,50M_\odot)$ \\
$\sigma_m$     & Width of the Gaussian component in the primary mass distribution & $U(1M_\odot,\,10M_\odot)$ \\
$\delta_m$     & Range of mass tapering at the lower end of the mass distribution & $U(0M_\odot,\,10M_\odot)$ \\
$\kappa$       & Power-law index of the BBH merger rate evolution with redshift & $U(-10,\,10)$ \\
\hline\hline
\end{tabular}
\label{tab:PLPeak_params_prior}
\end{table*}

\clearpage

\bibliographystyle{apsrev4-1-etal}
\bibliography{main}
\end{document}